\documentclass[prx,aps,superscriptaddress,reprint]{revtex4-2}

\usepackage{graphicx}
\usepackage{multirow}
\usepackage{amsmath,amssymb,amsfonts}
\usepackage{amsthm}
\usepackage{mathrsfs}
\usepackage{bm}
\usepackage{xcolor}
\usepackage{textcomp}
\usepackage{booktabs}
\begin{document}

\title{Learning Multiscale Dynamics of Nanoparticle Self-Assembly}

\author{Shuaifeng Li}
\affiliation{Department of Physics, University of Michigan, Ann Arbor, Michigan 48109, USA}
\author{Puquan Pan}
\affiliation{Department of Materials Science and Engineering, University of Illinois Urbana-Champaign, Urbana, Illinois 61801, USA}
\affiliation{Department of Electrical and Computer Engineering, University of Southern California, Los Angeles, California 90089, USA}
\author{Jonas Hallstrom}
\affiliation{Department of Physics, University of Michigan, Ann Arbor, Michigan 48109, USA}
\author{Nicholas Kotov}
\affiliation{Department of Chemical Engineering, University of Michigan, Ann Arbor, Michigan 48109, USA}
\author{Paul Bogdan}
\affiliation{Department of Electrical and Computer Engineering, University of Southern California, Los Angeles, California 90089, USA}
\author{Qian Chen}
\affiliation{Department of Materials Science and Engineering, University of Illinois Urbana-Champaign, Urbana, Illinois 61801, USA}
\author{Xiaoming Mao}
\email{maox@umich.edu}
\affiliation{Department of Physics, University of Michigan, Ann Arbor, Michigan 48109, USA}

\begin{abstract}
Nanoparticle self-assembly offers a powerful route to programmable functional materials, but its dynamics remain difficult to understand because they arise from multiscale, many-body, and environment-dependent interactions. Liquid-phase transmission electron microscopy~(LPTEM) enables real-time observation of these dynamics, yet extracting the underlying physical laws from highly stochastic trajectories remains a significant challenge. Here we develop a physics-informed, time-varying autoregressive framework that learns multiscale self-assembly dynamics directly from LPTEM videos. Continuous Eulerian fields with local density and orientational order are coarse-grained from individual video frames and evolved by a sparse, nearest-neighbor operator optimized over a constrained matrix manifold. This interpretable approach captures long-time large scale phase transitions by tracking operator volatility over time, and short-timescale local evolution and transport rules. Crucially, we extract the underlying partial differential equations, where expected symmetries naturally emerge, enabling the direct inference of fundamental coefficients. Our framework provides a robust, computationally efficient tool for discovering hidden physical mechanisms from observational data, opening a path toward the predictive design of self-assembled nanostructures.
\end{abstract}


\maketitle
\section{Introduction}\label{sec1}
Nanoparticle self-assembly offers a powerful route to create architected functional materials from nanoscale building blocks. Yet, given their length scales---intermediate between atoms governed by quantum mechanics and micron-scale colloidal particles in the classical regime---their dynamics is a complex product of van der Waals attraction, electrostatic repulsion, ligand-mediated forces, solvent effects, and entropic contributions~\cite{kotov1996ultrathin,lee2018mechanistic,whitesides2002self,silvera2015nonadditivity,boles2016self,ou2020kinetic}. These interactions can be anisotropic, many-body, and strongly dependent on the local environment, giving rise to stochastic rearrangements, transient clusters, kinetic arrest, and collective reorganization beyond conventional crystal-growth theories. Liquid-phase transmission electron microscopy~(LPTEM) is transforming this field by enabling real-time imaging of nanoparticle motion in solution with nanometer spatial and millisecond temporal resolution~\cite{ross2015opportunities} of collective nanoparticle dynamics. As a result, nanoparticle assemblies can now be directly visualized across diverse crystalline phases~\cite{chen2015interaction,kim2017imaging,luo2017quantifying,chen2024nanoparticle,qian2025nanoscale,lee2019real}.

However, visual access alone is not yet physical understanding. Nanoparticles motion in LPTEM exhibits  spatiotemporal heterogeneity, anomalous diffusion, and are governed by hierarchical interactions across multiple scales~\cite{jamali2021anomalous,weigel2011ergodic,kang2021real,vitali2018langevin}. Conventional analytical approaches, such as Langevin equations and generalized Langevin equations with memory kernels, are often unavailable in closed form or computationally prohibitive in heterogeneous environments~\cite{berkowitz1981memory,fricks2009time,chen2014computation,li2022spatiotemporal}. A central challenge is therefore to convert stochastic experimental trajectories into interpretable rules of nanoparticles dynamics.

Machine learning has provided powerful tools for analyzing such complex nanoparticle systems. Deep learning architectures such as U-Net have enabled automated particle segmentation and tracking in LPTEM videos~\cite{yao2020machine}. They have also been trained to learn effective inter-particle potentials directly from simulation trajectories and to predict the resulting self-assembled morphologies~\cite{gautham2022deep}, while unsupervised learning techniques have identified hidden nucleation pathways from experimental observations~\cite{adorf2019analysis}. More recently, generative models have combined attention-based transformer architectures with physics-informed loss functions to reproduce stochastic diffusion of nanoparticles in LPTEM~\cite{shabeeb2025learning}. These approaches are effective for pattern recognition, surrogate modeling, and trajectory generation, but their learned representations are typically implicit, and do not directly reveal explicit physics laws.

In contrast, autoregressive and operator-theoretic methods provide a complementary route that prioritizes interpretability over purely predictive accuracy. These methods, such as dynamic mode decomposition~(DMD), fit an explicit operator that advances the system state in time, extracts coherent spatiotemporal modes from high-dimensional time-series data, and finds broad applications in fluid dynamics, structural dynamics, and biological science~\cite{kutz2016dynamic,schmid2010dynamic,tu2014dynamic,li2023geometry,li2023characterization,li2023data,baddoo2023physics}. 

Challenges remain in applying these operator-theoretic methods to LPTEM videos. First, the dynamics are strongly time-varying. As the system evolves from a mobile disordered phase to an immobile ordered phase, even though the microscopic interactions among the NPs stay the same, the effective coarse-grained dynamics change continuously. Time-varying extensions of DMD can partially address this issue by allowing the operator to evolve in time~\cite{zhang2019online,dylewsky2022stochastically}. However, care must be taken in combining temporal adaptability with structural constraints to ensure that the results are physically meaningful~\cite{baddoo2023physics}. Second, to fully interpret the physics, a bridge is needed between discrete data-driven operators and continuum theory. Nanoparticle self-assembly systems are often governed by coupled fields. A framework needs to be developed to simultaneously handle the inter-field coupling, allow the governing equations to evolve continuously in time, and extract the full tensor of cross-field interaction and transport coefficients.

To address these challenges, we develop a physics-informed, time-varying autoregressive framework that learns interpretable local dynamical operators from LPTEM videos and reveals simple organizing principles underlying apparently complex nanoparticle self-assembly. This framework coarse-grains particle trajectories into Eulerian density $\rho$ and orientational $Q$ fields and learns a sequence of operators $A_{t}$ to linearly advance them in time such that $\begin{bmatrix}
    \rho_{t+1}\\Q_{t+1}
\end{bmatrix}=A_{t}\begin{bmatrix}
    \rho_{t}\\Q_{t}
\end{bmatrix}$. Each learned operator is constrained to a sparse nearest-neighbor structure and optimized on a matrix manifold, enforcing local physical interactions while reducing overfitting and preserving interpretability.

The learned dynamics naturally exhibit a low-rank and approximately reciprocal structure, consistent with equilibrium dynamics governed by local interactions, even though the system level dynamics appear heterogeneous and complex. It identifies the large scale phase transition where the system switches from disordered mobile state to a phase-separated state, consistent with the transition point identified through other graph-theory based structural descriptors~\cite{hallstrom2026decoding}. Finally, the discrete operators exhibit a convenient mapping to continuum partial-differential equations (PDEs). The resulting structure of the PDE is consistent with symmetry-based field theoretic expectations, even though these symmetries were not enforced in the learning algorithm.  Therefore, our methodology connects noisy experimental observations, interpretable data-driven operators, and continuum field theories of nanoparticle self-assembly within one unified framework, offering a consistent scope to extract simple rules underlying complex dynamics, and use them to steer assembly toward desired outcomes. 

\section{Results}\label{sec2}
\subsection{Sparse Autoregressive Modeling}\label{sec2_1}
Our goal is to develop an interpretable framework for learning complex, time-varying dynamics directly from observations of self-assembling systems. Such a framework should be able to capture dynamics that evolve across multiple spatial and temporal scales, while remaining sufficiently interpretable to connect the learned rules to underlying physical mechanisms. To test this capability, we focus on the self-assembly of corner-truncated gold nanocubes observed through LPTEM. This system has been reported to exhibit rich assembly dynamics~\cite{hallstrom2026decoding} from a dispersed phase with highly mobile nanocubes to a network of mesocrystals, which eventually merge into larger crystals~\cite{nagaoka2024dynamic}. Furthermore, the nanocubes are highly anisotropic, with both orientational and positional order arising during the assembly process, offering additional complexity beyond typical learning approaches to self-assembly involving spherical or isotropically interacting particles~\cite{adorf2019analysis,Dijkstra_2021}. Together, these characteristics make this system an informative platform for learning complex self-assembly dynamics.

We use two independent LPTEM experiments reported in Ref.~\cite{hallstrom2026decoding} as the primary experimental realizations of this system and analyze them here from a dynamical and continuum-field perspective. To complement the experimental observations, we additionally perform a molecular dynamics~(MD) simulation of nanoparticle self-assembly~\cite{zhou2022tunable}. The simulation provides a controlled realization of the assembly process in which the particle interactions and boundary conditions are explicitly specified, allowing us to examine how the same learning framework captures self-assembly dynamics across experimental and simulated settings.

The experimental nanocubes have an edge length of $63.3\pm5.2~\mathrm{nm}$ and an edge truncation of $17.1\pm1.2~\mathrm{nm}$ and are functionalized with carboxylate-terminated thiol ligands, which stabilize the particles through long-range electrostatic repulsion in water. Self-assembly is initiated by introducing phosphate-buffered saline~(PBS, $\mathrm{pH}=8$, ionic strength $I=15-26~\mathrm{mM}$) into the liquid cell, which screens the electrostatic repulsion and allows van der Waals attraction to drive the particles toward rhombic crystalline assemblies~\cite{ou2020kinetic,Luo_2023,Ou_2020}. LPTEM directly resolves the resulting evolution from a dispersed state through transient mesocrystals to extended crystalline domains at the single-particle level. Particle positions and in-plane orientations are extracted from each frame using a U-Net-based segmentation and tracking workflow~\cite{yao2020machine}. Full details of nanoparticle preparation, LPTEM imaging, and particle tracking are provided in Supplementary Note 1~\cite{supplementary}.
\begin{figure*}[t]
    \centering
    \includegraphics[width=0.8\textwidth]{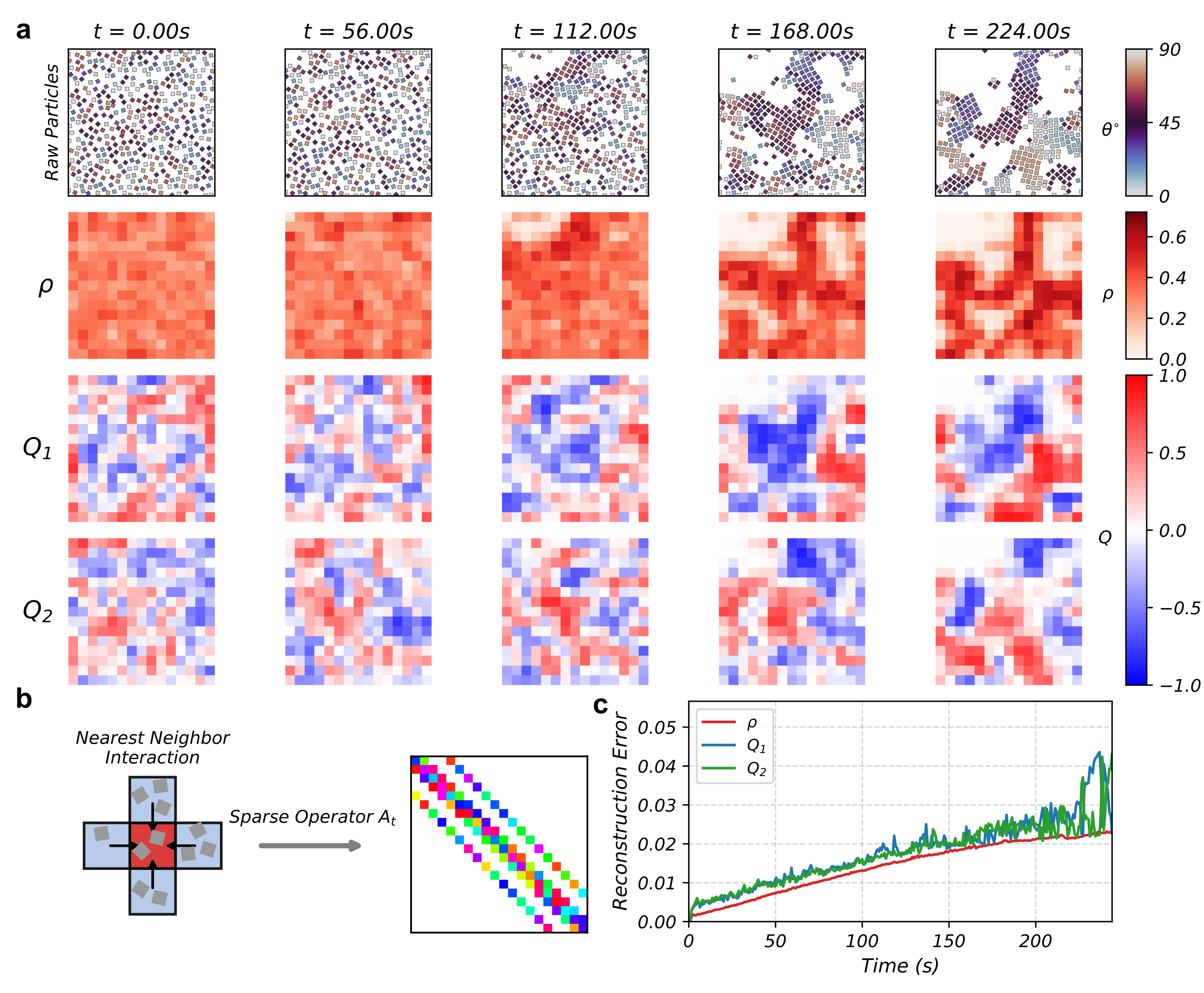}
    \caption{\textbf{Time-varying autoregressive modeling of nanoparticle self-assembly from LPTEM data.}
    \textbf{a} From top row to bottom row, representative snapshots of the raw particle positions colored by orientation angle $\theta$ and the corresponding coarse-grained continuous Eulerian fields,  $\rho$, $Q_{1}$ and $Q_{2}$, at five time points~($t = 0$, $56$, $112$, $168$, and $224~\mathrm{s}$) are shown in patches.
    \textbf{b} Schematic of the nearest-neighbor interaction constraint, resulting in a block-banded operator $A_{t}$.
    \textbf{c} Reconstruction error as a function of time for $\rho$, $Q_{1}$ and $Q_{2}$.
    }
    \label{fig1}
\end{figure*}

To bridge the discrete, stochastic particle trajectories imaged by LPTEM with both  operator learning and statistical mechanics continuum descriptions, we coarse-grain the data into continuous Eulerian fields on a regular spatial grid. The experimental field of view is partitioned into $N=N_{x}\times N_{y}$ spatial patches, and each frame across $T$ sequential time steps is processed by assigning particles to their patch and computing local statistics. The number of patches is chosen so that each patch contains enough particles to support a continuum coarse-grained description, while still retaining the spatial heterogeneity of the assembly. Hence, we choose $N_{x}=N_{y}=15$ in our case, as detailed in Supplementary Note 2 and Supplementary Figure 1~\cite{supplementary}. For cube nanoparticles in the field of view, two types of macroscopic fields are needed, one capturing the mass distribution and another capturing structural alignment. The local density field $\rho$ is defined as the normalized particle count per patch. To characterize the orientational alignment, we exploit the four-fold rotational symmetry of the cube nanoparticles. Individual particle angle $\theta$ is mapped to tetratic orientational order parameters $Q_{1} = \mathrm{Re}\langle e^{i4\theta} \rangle$ and $Q_{2} = \mathrm{Im}\langle e^{i4\theta} \rangle$, where the factor of $4$ reflects the equivalence of a square under $90^{\circ}$ rotations~\cite{steinhardt1983bond}. Together, these three fields~($X = [\rho, Q_1, Q_2]^T$, where each entry is an $N$-dimensional spatial vector) constitute a complete characterization of the spatiotemporal state. Figure~\ref{fig1}a shows the raw particles and three corresponding fields at representative time points~($0.0~\mathrm{s}$, $56.0~\mathrm{s}$, $112.0~\mathrm{s}$, $168.0~\mathrm{s}$ and $224.0~\mathrm{s}$) spanning the self-assembly process.

To discover the underlying physical laws governing the evolution of these fields, we adapt a time-varying autoregressive~(TVAR) linear framework~\cite{hamilton2020time,harris2021time,chen2023discovering,chen2025dynamic} with the physical information of nearest neighbor interactions. At each time step, the system state is advanced by an operator $A_{t}$, such that $X_{t+1} = A_{t}X_{t}$. Because the state comprises three coupled fields, each operator $A_{t}$ is a $3\times 3$ block matrix of size $3N\times 3N$. This block structure has direct physical meaning. The diagonal blocks capture intra-field dynamics such as density self-diffusion and orientational relaxation, while the off-diagonal blocks encode inter-field coupling. For instance, how local orientational alignment drives spatial fluxes in density, or how density influences the orientational field.

We determine the optimal sparse operator at each time step by minimizing a regularized loss function, $\mathcal{L}_{t}$, which balances data fidelity against three regularization penalties:
\begin{equation}
    \begin{split}
    \label{equ1}
    \mathcal{L}_{t} = &\frac{1}{2} ||X_{t+1} - A_t X_t||_2^2 + \lambda_1 ||A_t||_F^2 + \lambda_2 ||A_t - A_{t-1}||_F^2 \\
    & + \lambda_3 \text{tr}(A_t^T L A_t),
    \end{split}
\end{equation}
The first term enforces data fidelity. The second ($\lambda_{1}$) penalizes the Frobenius norm of the operator, promoting numerical stability and preventing overfitting to noise. The third ($\lambda_{2}$) enforces temporal smoothness by penalizing abrupt changes between consecutive operators. Physically, this encodes the expectation that system state should evolve gradually rather than discontinuously. The fourth ($\lambda_{3}$) uses the graph Laplacian $L$ of the spatial grid to promote spatial consistency of the learned interaction rules. This reflects approximate invariance that neighboring patches should obey similar update rules. These regularization weights are selected via a Monte Carlo search that minimizes the total prediction error and maintains interpretability, which are shown in Supplementary note 2~\cite{supplementary}.

Crucially, to ensure that the learned operator reflects true physical transport rather than data fitting, we enforce a hard structural prior. Each block of $A_{t}$ is restricted to a sparse, nearest-neighbor banded structure~(Figure~\ref{fig1}b). This encodes the known physics that local interactions dominate, where the state of each patch at the next time step depends only on its own value and those of its immediate spatial neighbors. The set of all matrices satisfying this sparsity mask forms a constrained manifold $\mathcal{M}$~\cite{absil2008optimization,baddoo2023physics}, and the optimization is solved at each time step using Riemannian gradient descent on this manifold~\cite{boumal2014manopt}:
\begin{equation}
    \label{equ2}
    \min_{A_t \in \mathcal{M}} \mathcal{L}.
\end{equation}

Solving this constrained optimization across all time steps yields a sequence of sparse linear operators that collectively govern the self-assembly process. To validate the learned operators, we use them to reconstruct the physical fields from the system's initial state by sequential application: $\hat{X}_{2}=A_{1}X_{1}$, $\hat{X}_{3}=A_{2}\hat{X}_{2}$, and so on. Despite the manifold constraints and regularization, the cumulative reconstruction error over the entire assembly process remains low~(approximately $1.4\%$ relative error for the total prediction, defined as $\frac{\|\hat{X}-X\|_{F}}{\|X\|_{F}}$, where $\|\cdot\|_{F}$ is the Frobenius norm), confirming that the sparse operators capture the dominant spatiotemporal dynamics~(see Figure~\ref{fig1}c for the time-resolved error and Supplementary Video $1$ for a visual comparison)~\cite{supplementary}. Both datasets yield comparably low reconstruction errors and reproduce the following analysis with qualitative features including low-rank cumulative dynamics, physically interpretable PDE coefficients. The full validation results are presented in Supplementary Notes $3-4$ and Supplementary Figures $2-3$~\cite{supplementary}.

\subsection{Long-Timescale Dynamics and Collective Behaviors}\label{sec2_2}
With the sequence of validated time-varying operators $A_{t}$, we first analyze the system at the macroscopic scale to uncover long-timescale collective behaviors. In nanoparticle self-assembly, the system undergoes distinct structural phases, from a highly mobile disordered state to an immobile crystalline state. A central question is when this transition occurs, and how to detect it directly from the learned operators. We address this by tracking the operator volatility, defined as the Frobenius distance between consecutive operators, $\|A_{t+1} - A_t\|_{F}$. The physical intuition is straightforward. When the system is in the disordered fluid phase or the ordered crystalline phase, the state changes slowly, and the temporal smoothness regularization ensures that consecutive operators remain similar. A pronounced peak in operator volatility therefore signals a shift in the system state. 
\begin{figure*}[t]
    \centering
    \includegraphics[width=0.8\textwidth]{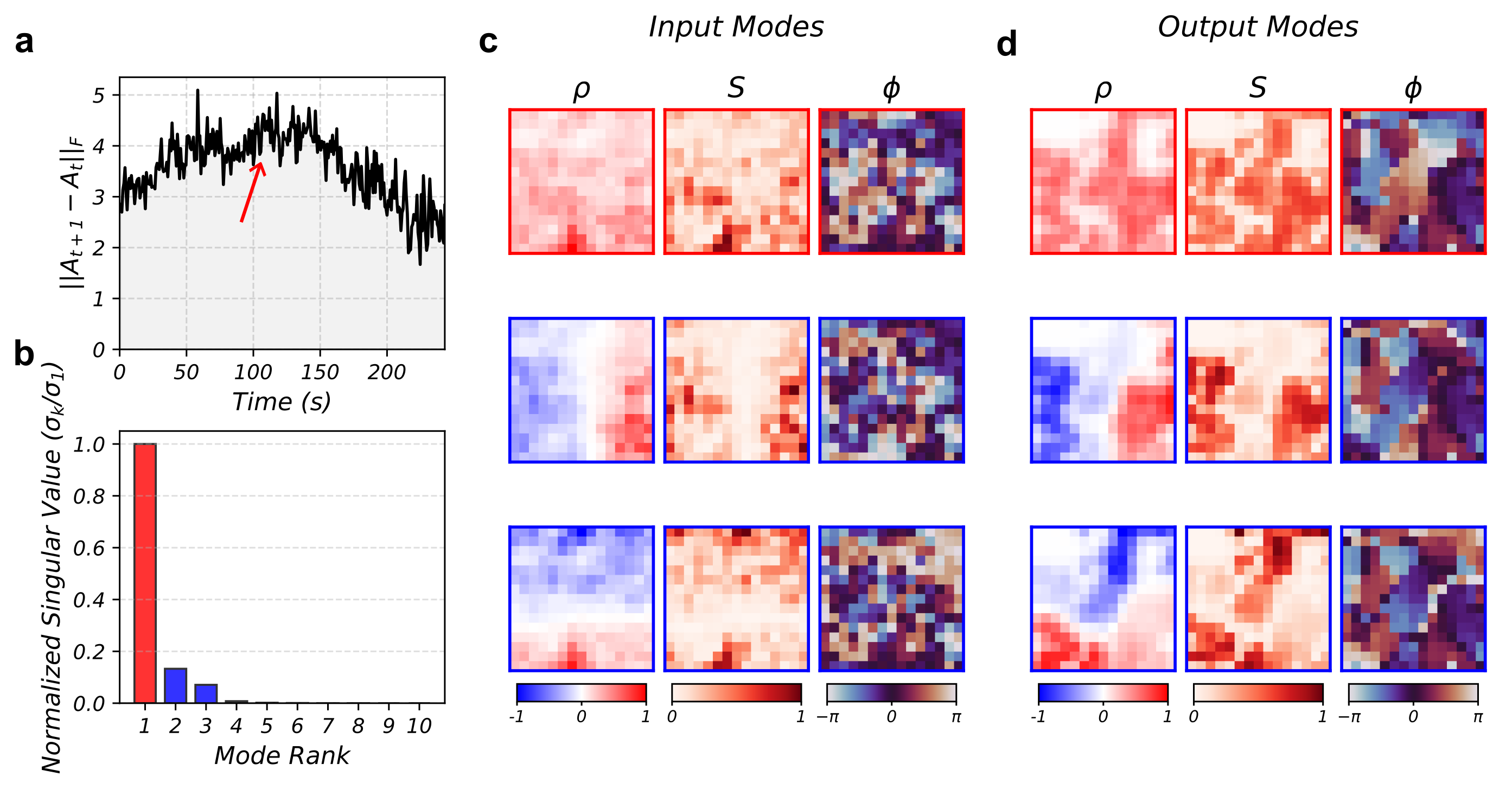}
    \caption{\textbf{Long-timescale collective dynamics revealed by spectral analysis of the cumulative propagator.}
    \textbf{a} Operator volatility $\|A_{t+1} - A_t\|_F$ as a function of time, with a peak indicated by a red arrow at around $110$ seconds signifying a switch in dynamics.
    \textbf{b} Normalized singular value spectrum of the cumulative propagator $\Phi = \prod_{t}^{T} A_{t}$, demonstrating that the long-timescale collective dynamics are dominated by only three modes.
    \textbf{c}, \textbf{d} Input and output spatial modes $1$ through $3$, displayed from top to bottom, in terms of density $\rho$, orientational order strength $S$, and phase angle $\phi$.
    }
    \label{fig2}
\end{figure*}

As shown in Figure~\ref{fig2}a, this metric reveals a clear and well-localized peak at approximately $110$ seconds into the assembly process. This timing is independently corroborated by an analysis of Ollivier–Ricci curvature of particle interaction networks, and is identified a minimum in this curvature at a closely matching time point~\cite{hallstrom2026decoding}. This agreement provides strong evidence that both methods are detecting the same underlying physical event, which is the critical reorganization from a disordered state to an ordered crystalline phase. We note that the operator volatility metric is sensitive to sharp transitions where the effective states change abruptly. In systems~(Experiment 2 and MD simulation) where the structural ordering proceeds gradually, as described in the Supplementary Notes $3-4$~\cite{supplementary}, the operator volatility remains approximately flat and this reflects the absence of the critical event. The presence or absence of a sharp operator variation peak thus can serve as a diagnostic that distinguishes abrupt phase transitions from continuous ordering processes.

Having identified when the system transitions, we next ask what structural patterns the system gravitates toward over the full trajectory. To answer this, we compute the cumulative time-evolution propagator $\Phi=\prod_{t=1}^{T}A_{t}$, which is the sequential product of all operators. This matrix maps the initial state directly to the final state: $X_{T}=\Phi X_{1}$. While each individual $A_{t}$ is sparse and encodes only nearest-neighbor interactions, their cumulative product $\Phi$ is dense and encodes the effective long-range spatiotemporal flow of the entire assembly process. Singular value decomposition of $\Phi = U\Sigma V^*$ then decomposes this global map into a hierarchy of dominant spatial modes, ranked by their amplification over the trajectory. Here, the right singular vectors $V$ represent input spatial modes showing initial density-orientation patterns that are preferentially propagated by the cumulative dynamics. The left singular vectors $U$ represent the corresponding output spatial modes indicating the density-orientation patterns result from those inputs. The rapid decay of the singular value spectrum in Figure~\ref{fig2}b reveals that, despite the apparently complex and disordered nature of the raw trajectories, the long-timescale collective dynamics are dominated by only three modes, representing the low-rank structure. The instantaneous operators $A_{t}$ may have a richer spectral structure at any given time step, but the cumulative effect of the full assembly process concentrates the dynamics into a small number of structural attractors.

To physically interpret these modes, particularly the orientational alignment, we convert $Q_{1}$ and $Q_{2}$ into polar coordinates. This yields the orientational order strength, $S = ||Q_{1} + iQ_{2}||_2$, which measures the degree of local alignment regardless of direction, and the phase angle, $\phi = \text{arg}(Q_{1} + iQ_{2})$, which specifies the preferred orientation axis. In Figure~\ref{fig2}c, the leading input mode~(mode $1$), obtained from the right singular vectors of $\Phi$, exhibits monopole characteristics in both $\rho$ and $S$. We observe a spatially uniform pattern representing bulk accumulation and the global growth of orientational order across the field of view. This mode captures the thermodynamic driving force toward crystallization.

In contrast, modes $2$ and $3$ exhibit dipole structures, where spatial patterns with one positive and one negative lobes are antisymmetric around the center of the field. These modes capture the symmetry-breaking spatial gradients that drive grain boundary formation between crystalline domains with different orientational alignment. The emergence of two dipole modes reflects the two independent spatial directions along which such gradients can develop. Together, these three modes provide a physically interpretable basis for the long-timescale self-assembly process, where a uniform crystallization drive~(monopole) is modulated by directional domain competition~(dipoles).

The output spatial modes obtained from the left singular vectors of $\Phi$, reveal the structural patterns that characterize the final state produced by the cumulative dynamics, as shown in Figure~\ref{fig2}d. Output mode $1$ differs markedly from its input counterpart, displaying a spatially heterogeneous pattern in $\rho$, with localized clusters of high orientational order in $S$ and sharply partitioned domains in $\phi$. This structure directly encodes the polycrystalline microstructure of the final state, where multiple crystalline grains own distinct orientational alignments, separated by grain boundaries. Output modes $2$ and $3$ exhibit more fragmented spatial features that resolve the specific locations and orientations of individual grain boundaries in the final assembly. The analysis shows how the cumulative propagator transforms the dynamics, where system-scale ordering tendencies encoded in $V$ are progressively refined through the sequence of operators into the polycrystalline microstructure encoded in $U$.

\subsection{Short-Timescale Dynamics and Interaction Patterns}\label{sec2_3}
The preceding analysis characterizes the global structures encoded by the cumulative propagator. To complement this long-timescale view, we next examine how local information is transported through this assembling system, which is an informative way to show underlying physics. Specifically, we ask how a localized perturbation in one patch affects neighboring and distant patches across the coupled fields. 

\begin{figure}[t]
    \centering
    \includegraphics[width=0.48\textwidth]{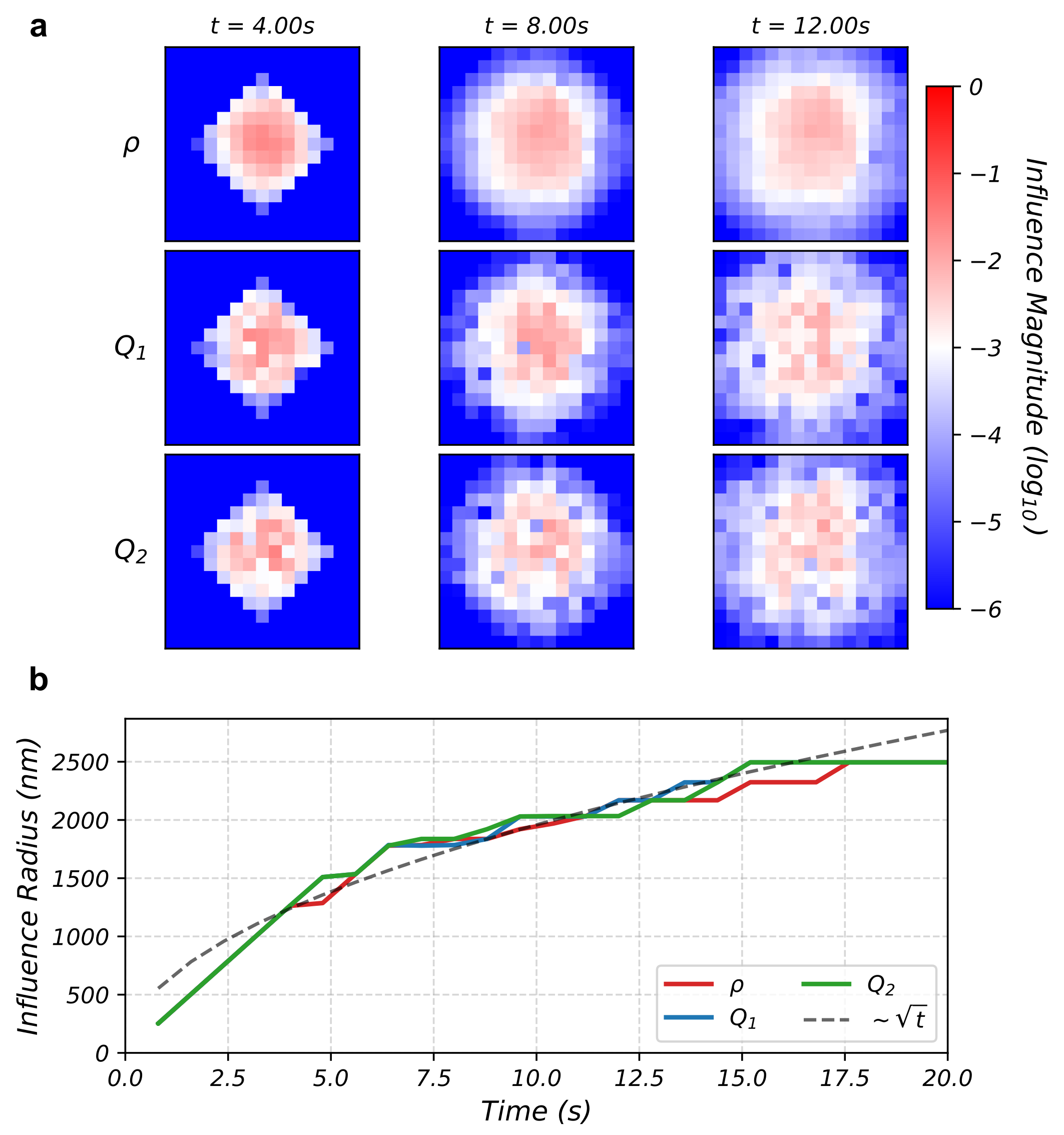}
    \caption{\textbf{Spatiotemporal impulse response of the nanoparticle self-assembly system.}
    \textbf{a} Spatial maps of the impulse response to a localized density perturbation at the center of the field, shown for all three fields at $t = 4.0$, $8.0$, and $12.0~\mathrm{s}$.
    \textbf{b} Influence radius as a function of time for each field, alongside the theoretical $R \sim \sqrt{t}$ scaling~(dashed gray line). 
    }
    \label{fig3}
\end{figure}

Since the learned operators define the linearized time evolution of these fields, this response can be computed directly from the partial propagator $\Phi(t_{0},t_{0}+\tau)=\prod_{s=t_{0}}^{t_{0}+\tau} A_{s}$, which maps the system state at time $t_{0}$ to the state $\tau$ steps later. Each column of $\Phi(t_{0},t_{0}+\tau)$ gives the response of the entire system to a unit perturbation applied to a specific field at a specific spatial patch, and therefore serves as the discrete Green's function of the learned dynamics. Extracting the column associated with a density perturbation at the center of the field and tracking its spatial structure at $\tau = 4.0~\mathrm{s}$, $8.0~\mathrm{s}$, and $12.0~\mathrm{s}$~(Figure~\ref{fig3}a) reveals the outward spread of influence from the perturbation source. The response is spatially localized at short times and progressively broadens, with the perturbation spreading into all three fields, indicating that a density fluctuation not only diffuses spatially but also couples into the orientational order fields $Q_{1}$ and $Q_{2}$.

To quantify the propagation rate, we define the influence radius $R$ as the maximum distance from the source at which the response exceeds a threshold magnitude, and track $R$ as a function of time for each output field~(Figure~\ref{fig3}b). The influence radius grows as $R\sim t^\alpha$ with a fitted exponent consistent with $\alpha\approx0.5$ for all three fields, confirming that information propagation in the nanoparticle system is fundamentally diffusive. The complete set of impulse responses for perturbations originating in the $Q_{1}$ and $Q_{2}$ fields is presented in Supplementary Note $5$~\cite{supplementary}, where the same diffusive scaling is observed for all source–output field combinations. This $\sqrt{t}$ scaling stands in contrast to the linear expansion $R\sim t$ that would characterize ballistic transport in a momentum-conserving system, and indicates that the inter-particle interactions in the liquid cell are dominated by dissipative dynamics, consistent with the viscous environment of the solution-phase liquid cell.

The Green's function analysis establishes that perturbations propagate diffusively through the assembling system, but diffusive transport can arise from many different microscopic coupling mechanisms acting simultaneously apart from liquid-cell environment. Therefore, it would be interesting to explore whether the complex interaction rules embedded in the learned operators can be decomposed into a small number of specific physical mechanisms, each with its own characteristic interaction pattern, spatial footprint, and temporal activation profile.

The nearest-neighbor sparsity constraint gives each block of $A_{t}$ the structure of a local spatial stencil. For a given input field $\beta$ and output field $\alpha$, the update $X_{\alpha}$ at a spatial patch depends only on $X_{\beta}$ at the same patch and at its four nearest neighbors. Hence, each local stencil contains five weights which are a center weight $w_{C}$ and four directional weights $w_{E}$, $w_{W}$, $w_{N}$, $w_{S}$ corresponding to the east, west, north and south neighboring patches. These weights specify the local evolution rule by determining how the input-field values in this five-patch neighborhood contribute to the output-field update at the center patch. Because the system state contains three fields, there are $3\times 3=9$ input-output stencil types at each spatial location and time point. For example, the density update can be written as $\rho_t = \rho_{t-1} * K_{\rho \to \rho,t} + Q_{1,t-1} * K_{Q_1 \to \rho,t} + Q_{2,t-1} * K_{Q_2 \to \rho,t}$, where each $K_{\beta\to\alpha,t}$ denotes the corresponding five-point local stencil and $*$ denotes the stencil operation. Collectively, the $9$ stencil types and their five weights give $9\times 5=45$ coupling weights per spatial patch and time step. We stack these weights into an interaction tensor $\mathcal{I} \in \mathbb{R}^{45 \times N \times T}$, whose three axes correspond to evolution type, spatial dimension and time dimension.
\begin{figure}[t]
    \centering
    \includegraphics[width=0.48\textwidth]{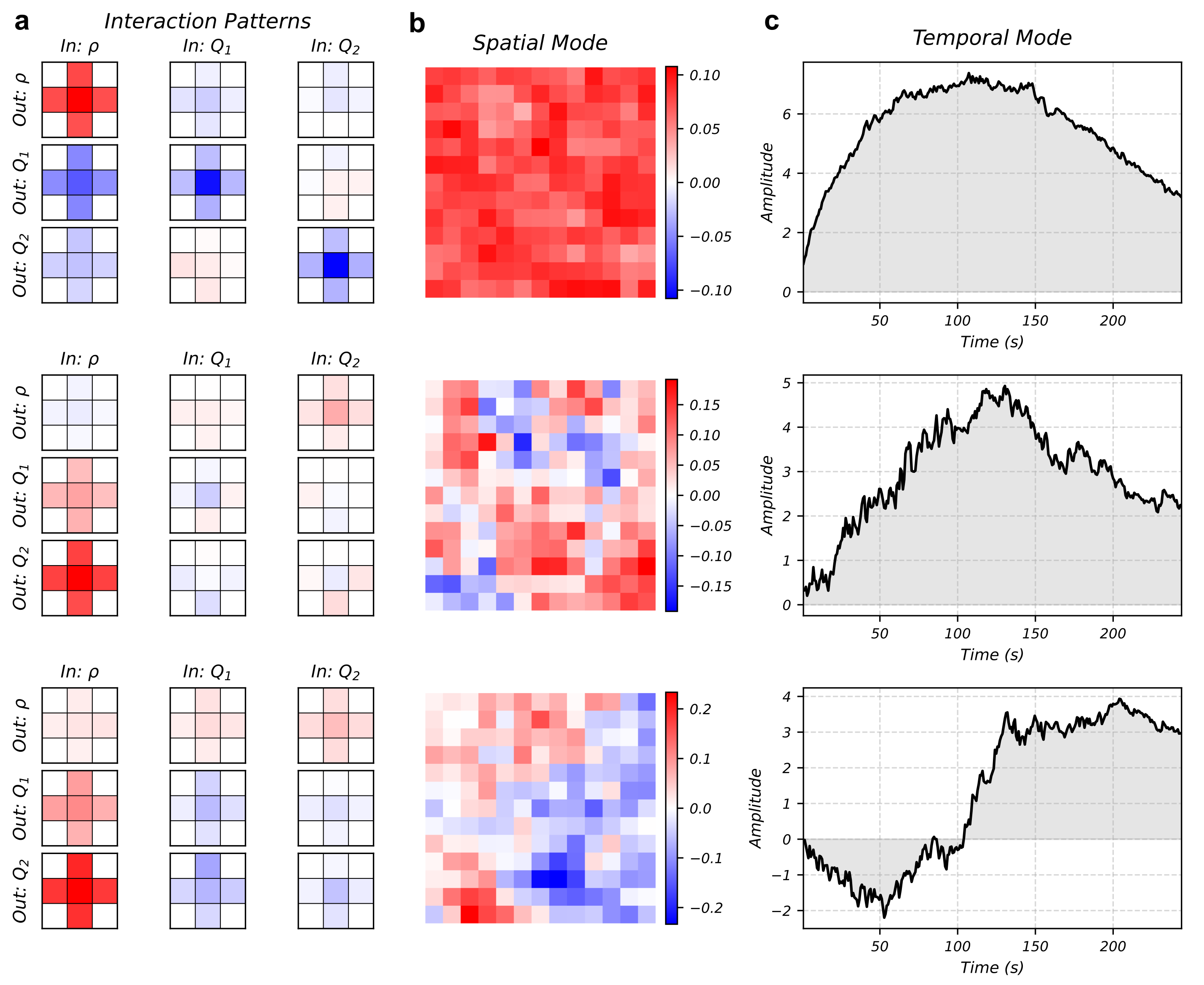}
    \caption{\textbf{Top three interaction patterns extracted from CP decomposition.}
    Each row presents one rank-one component extracted by CP decomposition of the interaction tensor $\mathcal{I} \in \mathbb{R}^{45 \times N \times T}$.
    \textbf{a} The $3\times3$ interaction stencils for all nine input-output field pairings ($\rho$, $Q_{1}$, $Q_{2}$), showing the normalized spatial coupling pattern for each component.
    \textbf{b} The spatial mode, indicating the macroscopic regions of the assembly where each interaction mechanism is most strongly active.
    \textbf{c} The temporal amplitude, showing when each mechanism is engaged during the self-assembly.
    }
    \label{fig4}
\end{figure}

To decode this high-dimensional and time-varying set of evolution rules, we apply Canonical Polyadic~(CP) decomposition to factorize it into a sum of rank-one components: $\mathcal{I} = \sum_{r=1}^R u_r \circ v_r \circ w_r$~\cite{kolda2009tensor}. CP decomposition separates three axes of variation, interaction type, spatial location, and time, into independent factors. Each component $r$ consists of three interpretable vectors. The evolution mode $u_{r}$ captures the normalized evolution pattern, where the specific shape of the $3\times3$ stencils across all $9$ input–output field pairings reveal which inter-field couplings dominate in that component. The spatial mode $v_{r}$~(dimension $N$) reveals where this coupling mechanism is most active. The temporal mode $w_{r}$~(dimension $T$) dictates when during the assembly trajectory this mechanism turns on or off.

Figure~\ref{fig4} displays top three components with the largest weights. Each row presents one component as three linked panels, the $3\times3$ grid of interaction stencils~(left), the spatial mode~(center), and the temporal amplitude~(right). The first component is dominated by the density column and the $\rho\to\rho$ stencil shows strong positive weights characteristic of local accumulation and self-diffusion, while the $\rho\to Q_{1}$ and $\rho\to Q_{2}$ stencils indicate a negative feedback of density on orientational order, where increasing local density suppresses orientational order. Couplings from $Q_{1}$ and $Q_{2}$ as input fields are negligible. The spatial mode is uniformly positive across the field of view, and the temporal mode rises steadily, peaks around $110~\mathrm{s}$, and then gradually declines. This component captures a global density accumulation mechanism in which crowding frustrates local orientational alignment, in consistent with the early disordered phase where particles are concentrated but not yet geometrically constrained into a preferred orientation.

The second component also draws primarily from density as the input field, but its effect is spatially heterogeneous. The $\rho\to Q_{1}$ and $\rho\to Q_{2}$ stencils indicate density-orientation coupling, while the spatial mode contains both positive and negative regions. Therefore, this mechanism promotes orientational order in regions where the spatial mode is positive and suppresses orientational order where spatial mode is negative. Its temporal mode peaks sharply during the phase transition window and then decays. This component captures a spatially localized crystallization mechanism in which local density drives the particles into aligned configurations, representing the onset of crystalline grain formation. The opposing signs of the density-to-orientation coupling in components $1$ and $2$ reveal a possible competition. Density can either frustrate or promote alignment depending on the local spatial context and the stage of assembly, and the CP decomposition is able to separate these two regimes into independent mechanisms.

The third component shares several features with the second component, including activity in the density-to-orientation coupling and a heterogeneous spatial mode, but it also shows stronger activity in the orientational diagonal blocks ($Q_{1}\to Q_{1}$ and $Q_{2}\to Q_{2}$). This indicates that this component captures not only density-to-orientational alignment, but also the subsequent redistribution of orientational order within the $Q_{1}$ and $Q_{2}$ fields. Its temporal mode changes sign during the assembly, being negative before the transition and positive afterward. This sign reversal suggests that the same coupling mechanism plays opposing physical roles in the two phases. From the spatial mode, regions where the component initially weakens orientational coupling later become regions where the same coupling reinforces crystalline alignment, vice versa.

Beyond the individual CP components, we also quantify how spatially homogeneous the learned evolution rules are across the field of view. For each spatial patch, we compute a stencil heterogeneity index $H(t)$, defined as the relative difference between neighboring local evolution rules. As shown in Supplementary Note 6 and Supplementary Figure 5~\cite{supplementary}, $H(t)$ gradually decreases during assembly, indicating that the learned local rules become more spatially homogeneous as the system approaches the final crystalline state and the assembly structure is getting more uniform in the motif. This trend supports the approximate translational-invariance assumption encoded by the graph-Laplacian regularization in the loss function, while retaining the spatial heterogeneity needed to resolve grains and grain boundaries.

A notable structural feature across all three components is the approximate symmetry of the interaction stencils along the block diagonal. The pattern coupling field $\alpha$ to field $\beta$ closely resembles the pattern coupling $\beta$ to $\alpha$. This symmetry is not imposed by the optimization as the manifold constraint only enforces spatial locality but places no restriction on the relationship between off-diagonal blocks. The emergent near-symmetry is the data-driven manifestation of Onsager reciprocal relations near thermodynamic equilibrium~\cite{onsager1931reciprocal1,onsager1931reciprocal2}. Onsager reciprocity is a consequence of microscopic reversibility and its appearance in the learned operators provides model-free evidence that the nanoparticle self-assembly proceeds as a near-equilibrium process.

\subsection{Inference of Continuum Theory and PDE Extraction}\label{sec2_4}
Our TVAR framework provides a direct bridge from the discrete learned operators to classical continuum theory. The TVAR model $X_{t+1} = A_{t}X_{t}$ advances the discrete fields by one time step. Rewriting this as a finite difference equation: $\frac{X_{t+1}-X_t}{\Delta t} = \frac{A_{t} - I}{\Delta t} X_t$ shows that the learned operator $\frac{A_{t}-I}{\Delta t}$ approximates the generation of the continuum dynamics. Because $A_{t}$ is organized into $3\times 3$ field-coupling blocks, this generator naturally yields a fully coupled reaction-advection-diffusion equation:
\begin{equation}\label{equ3}
  \begin{split}
    \partial_{t}X_{\alpha} = &\sum_{\beta}\left[D_{\beta\to\alpha}^{x}\partial_{x}^2X_{\beta}+D_{\beta\to\alpha}^{y}\partial_{y}^2X_{\beta}\right.\\
    &\left.+v_{\beta\to\alpha}^{x}\partial_{x}X_{\beta}+v_{\beta\to\alpha}^{y}\partial_{y}X_{\beta}\right.\\
    &\left.+R_{\beta\to\alpha}X_{\beta}\right],
  \end{split}
\end{equation}
where $\alpha,\beta\in\{\rho,Q_{1},Q_{2}\}$. The diagonal terms describe self-transport and self-relaxation of each field, while the off-diagonal terms describe cross-coupling, such as density-driven orientational ordering or orientational gradients feeding back into density evolution.

\begin{figure}[t]
    \centering
    \includegraphics[width=0.50\textwidth]{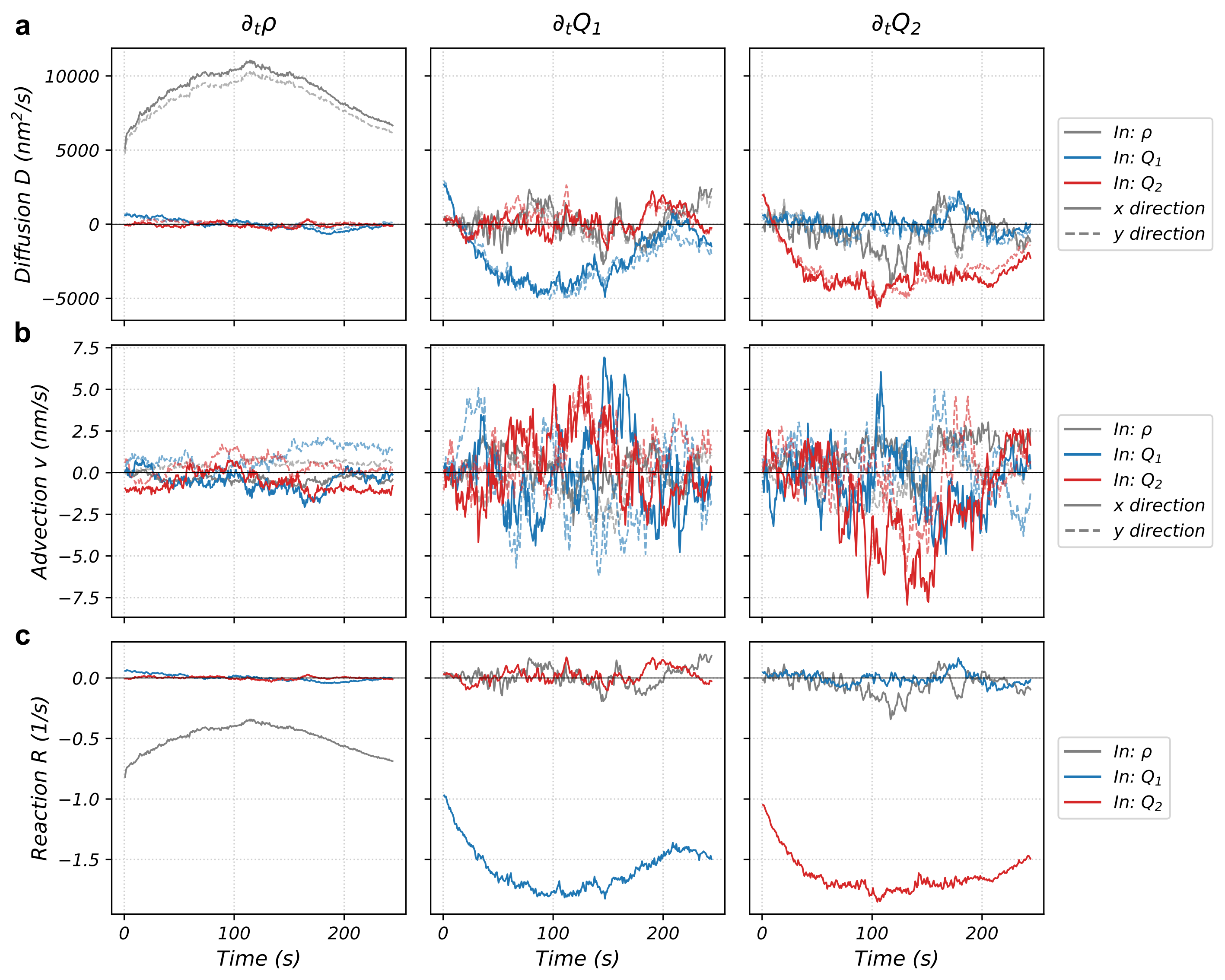}
    \caption{\textbf{Extraction of continuum PDE coefficients from spatial moments of the learned operators.}
    Time-resolved PDE coefficients recovered from the spatial moments of the learned $3\times3$ evolution stencils and averaged over spatial patches, organized by output field (columns: $\partial_t\rho$, $\partial_t Q_1$, $\partial_t Q_2$) and coefficient type (rows: diffusion $D$, advection $v$, reaction $R$). Line color denotes the input field ($\rho$: gray, $Q_1$: blue, $Q_2$: red); solid and dashed lines distinguish the $x$ and $y$ spatial directions for $D$ and $v$. 
    \textbf{a} The density self-diffusion is positive and grows steadily before declining as clusters consolidate. In the orientational order equations ($\partial_t Q_1$, $\partial_t Q_2$), the self-diffusion coefficients  are negative. 
    \textbf{b} Advection coefficients fluctuate around zero across all fields. 
    \textbf{c} Self-reaction terms for all fields are negative, acting as a stabilizing decay that bounds field magnitudes.
    }
    \label{fig5}
\end{figure}

The PDE coefficients are extracted from the spatial moments of the learned local stencils. Each block of $\frac{A_{t}-I}{\Delta t}$ acts as a local five-point stencil mapping an input field $\beta$ to an output field $\alpha$, with center, east, west, north, and south weights denoted by $w_{C}$, $w_{E}$, $w_{W}$, $w_{N}$ and $w_{S}$. A Taylor expansion of the stencil gives the leading continuum coefficients: $D_{\beta\to\alpha}^{x}=\frac{w_{E}+w_{W}}{2}\frac{\Delta x^2}{\Delta t}$, $v_{\beta\to\alpha}^{x}=\frac{w_{W}-w_{W}}{2}\frac{\Delta x}{\Delta t}$, with an analogous expression for $D_{\beta\to\alpha}^{y}$ and $v_{\beta\to\alpha}^{y}$, $R = \frac{w_C + w_N + w_S + w_E + w_W - \delta}{\Delta t}$, where $\delta=1$ for diagonal blocks and $\delta=0$ for off-diagonal blocks. At each time step, these moment relations are first applied to the local stencil at each spatial patch corresponding to all nine input-output field, yielding coefficients for each patch. The diffusion, advection, and reaction coefficients shown in Figure~\ref{fig5} are then obtained by averaging these patch-dependent values over the field of view. This method should be considered as an effective continuum closure because it assumes that the coarse-grained fields vary smoothly over a grid spacing and hence gives spatially averaged transport coefficients.

The extracted coefficients can be interpreted following a Ginzburg-Landau (GL) picture for coupled fields~\cite{allen1979microscopic,landau1937theory,hohenberg1977theory,korbuly2017orientation,granasy2004modelling}. In this picture, a phenomenological free-energy density can be written as:
\begin{align}
    f(\rho,\mathbf{Q})=f_{\rho}(\rho) + f_{\mathbf{Q}}(\mathbf{Q}) + f_c(\rho,\mathbf{Q}).
\end{align}
We can expand this free energy to leading order of concentration fluctuations $\tilde{\rho}\equiv \rho -\rho_0$ (where $\rho_0$ is the mean concentration) and orientational order $\mathbf{Q}$, 
\begin{align}
    f_{\rho}(\rho) &= f_{\rho}(\rho_0) + \frac{a}{2}\tilde{\rho}^{2} +\frac{\kappa_{\rho}}{2}|\nabla\tilde{\rho}|^{2}, \nonumber\\
    f_{\mathbf{Q}}(\mathbf{Q}) &=\frac{A}{2}S^{2} +\frac{1}{2}\mathrm{Tr}[(\nabla\mathbf{Q})^{T}\mathbf{\kappa_{Q}}(\nabla\mathbf{Q})], \nonumber\\
    f_c(\rho,\mathbf{Q})&=g\tilde{\rho} S^2,
\end{align}
where we have used the internal symmetries of the fields, external rotational symmetry, and defined $S^{2}=Q_{1}^{2}+Q_{2}^{2}$.  The $\mathbf{\kappa_{Q}}$ term contracts the sub-indices of $\mathbf{Q}$ and the spatial derivatives separately. Written explicitly, this contraction is $\frac{1}{2}\sum_{\mu,\nu=x,y}\kappa_{Q,\mu\nu}\sum_{i,j=x,y}(\partial_{\mu}Q_{ij})(\partial_{\nu}Q_{ji})$, where $\mu$, $\nu$ denote spatial derivative directions and $i$, $j$ denote orientational tensor indices. The form of the coupling term is because although $\mathbf{Q}$ obeys rotational symmetry of the embedding space, its coupling to spatial derivatives starts at the fourth order, given its four-fold rotational symmetry. The leading order coupling allowed by symmetry is $\tilde{\rho} S^2$, which is beyond quadratic.

The phase separation and crystallization that we observe correspond to a bistable form of $f_{\rho}(\rho)$, with two minima at $\rho_L$ and $\rho_H$, corresponding to low- and high-density states, separated by an unstable intermediate density $\rho_M$. At early times, the density field is relatively homogeneous, whereas during assembly it becomes increasingly heterogeneous, with local densities evolving toward the low- and high-density states. As a result, the coefficients $a,\kappa_{\rho},A,\mathbf{\kappa_Q}$ all depend on $\rho_{0}$, evolving over time and becoming increasingly heterogeneous in space. 

Using this free energy, we can then write down the time-dependent Allen-Cahn dynamics~\cite{allen1979microscopic}:
\begin{equation}
    \label{equ6}
    \partial_{t}\tilde{\rho}=-\Gamma_{\rho}\frac{\delta\mathcal{F}}{\delta\tilde{\rho}}, \quad
    \partial_{t}\mathbf{Q}=-\Gamma_{\mathbf{Q}}\frac{\delta\mathcal{F}}{\delta\mathbf{Q}},
\end{equation}
with kinetic matrix $\Gamma_{\rho}$ and $\Gamma_{\mathbf{Q}}$. Here $\mathcal{F}[\rho,\mathbf{Q}]\equiv\int d^{2}\mathbf{r} f(\rho,\mathbf{Q})$. This gives
\begin{equation}
    \label{equ7}
    \partial_{t}\tilde{\rho}=D_{\rho}\nabla^{2}\tilde{\rho}-\Gamma_{\rho}a\tilde{\rho} ,
\end{equation}
\begin{equation}
    \label{equ8}
    \partial_{t}\mathbf{Q}=\mathbf{D_{Q}}\nabla^{2}\mathbf{Q}-\mathbf{\Gamma_{Q}}A\mathbf{Q},
\end{equation}
where $D_{\rho}=\Gamma_{\rho}\kappa_{\rho}$, $\mathbf{D_{\mathbf{Q}}}=\Gamma_{\mathbf{{Q}}}\mathbf{\kappa_{Q}}$. The full derivation is provided in Supplementary Note 7~\cite{supplementary}. Note that we use non-conserved dynamics for $\rho$ as the LPTEM field of view is an open system.  This symmetry based analysis predicts that in the linear order, the PDE system of $\tilde{\rho},\mathbf{Q}$ has no cross-coupling between the $\tilde{\rho},\mathbf{Q}$ fields, but the coefficients depend on the $\rho_0$ we expand from.

We then analyze our learned PDEs with the guidance from this GL framework. In particular, given the limitations of the system size, we take the total concentration field $\rho$ instead of using $\tilde{\rho}=\rho-\rho_0$ with $\rho_0$ being a spatially heterogeneous for our discussion of phase separation, but the basic phenomena can still be understood in the GL framework.

Figure~\ref{fig5}a shows the self-diffusion coefficients $D_{\rho\to\rho}$ for Experiment $1$. This term, defined theoretically as $D_{\rho}=\Gamma_{\rho}\kappa_{\rho}$, is consistently positive-definite. It indicates ordinary diffusive relaxation where local density inhomogeneities are smoothed as particles redistribute. This coefficient first increases during the highly mobile stage and then declines as the crystalline clusters consolidate and particles are less mobile.

In contrast, the orientational fields exhibit fundamentally different behavior. The extracted self-diffusion coefficients $D_{Q_{1}\to Q_{1}}$ and $D_{Q_{2}\to Q_{2}}$ have negative spatially-averaged values throughout the assembly trajectory. Because the Cartesian components $Q_{1}$ and $Q_{2}$ are not individually rotationally invariant, we validate this instability by confirming that the eigenvalues of the full $2\times2$ orientational diffusion matrix $\mathbf{D_{Q}}$ are also negative~(Supplementary Note 7)~\cite{supplementary}. In the continuum theory, where $\mathbf{D_{Q}}=\Gamma_{\mathbf{Q}}\mathbf{\kappa_{Q}}$, a positive kinetic factor $\Gamma_{Q}$ necessitates a negative spatial stiffness $\kappa_{Q}$. This negative orientational diffusion can be understood using the GL framework, where all coefficients are functions of $\rho_0$. At interfaces between low and high $\rho_0$ phases, the abrupt change of $\rho_0$ require higher order coupling terms to be included, which can change the sign of $\kappa_{Q}$. Indeed, the self-diffusion coefficient exhibit strong spatial heterogeneities~(Supplementary Note 8 and Supplementary Figure 7), consistent with this view~\cite{supplementary}. 

The advection coefficients fluctuate around zero across all fields shown in Figure~\ref{fig5}b, consistent with the absence of a persistent imposed flow in the liquid cell. The potentially unbounded growth driven by the anti-diffusive orientational instability is counterbalanced by local reaction terms. Figure~\ref{fig5}c demonstrates that the self-reaction coefficients $R$ for all fields are negative. In the Ginzburg-Landau framework, these correspond to the stabilizing local free-energy penalties, such as $-\Gamma_{Q}AQ$. Thus, the nanoparticle ordering process is revealed to be a direct competition, where anti-diffusive amplification drives the sharpening of orientational domains, while local reaction terms provide stabilizing decay to bound the magnitudes. 

Finally, while the off-diagonal diffusion coefficients are smaller than the dominant self-diffusion terms, they encode physical cross-talk, mapping how local density gradients actively influence orientational ordering and how existing orientational structure feeds back into density redistribution.

In the second LPTEM experiment and the MD simulation, the orientational diffusion eigenvalues are strictly positive. In the context of the coarse-grained free energy, this corresponds to a positive orientational spatial stiffness, placing these systems in a standard diffusive relaxation regime. Hence, orientational gradients are progressively smoothed, leading to a gradual ordering process. This behavior stands in stark contrast to the anti-diffusive instability of Experiment $1$, where the negative eigenvalues drive abrupt ordering that coincides with the pronounced peak in macroscopic operator volatility~(Figure~\ref{fig2}a). Across all three datasets, the stabilizing local reactions and positive density diffusion trends remain qualitatively universal. It is exclusively the sign of the orientational diffusion eigenvalues that dictates whether the system undergoes anti-diffusive spatial domain sharpening or ordinary diffusive relaxation. Ultimately, this moment-based PDE extraction can not only recover effective continuum transport coefficients but also provides a data-driven method to distinguish the distinct kinetic pathways of nanoparticle crystallization.

\section{Discussion}
In this work, we have presented a data-driven, physics-informed framework for discovering the multiscale dynamics governing nanoparticle self-assembly directly from experimental LPTEM videos, which is also complemented by MD simulation. By coarse-graining discrete particle trajectories into continuous Eulerian fields and fitting a time-varying autoregressive model on a sparse manifold, we obtain a sequence of interpretable linear operators that reconstruct the full assembly dynamics with a cumulative error of approximately $1.4\%$. The key design choice, which is enforcing nearest-neighbor locality as a hard structural constraint via manifold optimization, bypasses the opacity of black-box models and yields operators whose internal structure can be directly interpreted at multiple scales.

The framework extracts a coherent hierarchy of physical insight from a single learned operator sequence, without requiring separate models and analysis pipelines for each scale. At the macroscopic level, operator volatility provides a model-free indicator of phase transitions. Spectral decomposition of the cumulative propagator reveals that the long-timescale ordering is governed by a small number of structural modes with clear physical identities, consistent with near-equilibrium dynamics governed by local rules. At the mesoscopic scale, the discrete Green's function maps diffusive propagation of local perturbations with quantitative $\sqrt{t}$ scaling across all field couplings. Tensor decomposition of the learned stencils isolates distinct interaction mechanisms and resolves their spatial and temporal activation profiles. Finally, spatial moment analysis of the stencils recovers the underlying PDEs. Remarkably, even though symmetries were not enforced in our learning method, they naturally emerge in a structure as expected from GL theory. Importantly, different realizations in another experiment and MD simulation are not expected to exhibit identical dynamics. Rather, they provide complementary examples through which the framework can identify and distinguish different dynamical pathways of self-assembly.

Several limitations of the current framework should be noted. First, the autoregressive model is linear at each time step. While linearity is a reasonable approximation for coarse-grained fields evolving over short time intervals, strongly nonlinear phenomena may not be fully captured. Extending the framework to include weakly nonlinear corrections, for instance, through polynomial or kernel-based generalizations of the operator, is a natural next step. Second, the nearest-neighbor interaction constraint assumes that the dominant couplings are short-ranged. For systems with significant long-range interactions, the interaction radius would need to be increased, at the cost of a larger parameter space. Third, the moment-based PDE extraction relies on the coarse-grained fields varying smoothly over a grid spacing, which is an assumption that may break down near sharp grain boundaries. The extracted PDE coefficients should therefore be interpreted as spatially averaged, effective transport parameters rather than exact values.

It is instructive to compare our approach to related methods. The statistical PDE learning framework demonstrated that reaction-convection-diffusion models can be inferred from in-situ STEM videos~\cite{li2020statistical}, but assumed time-invariant coefficients and a single scalar field, requiring manual temporal segmentation when the dynamics change regime. Our time-varying, multifield formulation handles these transitions automatically and extracts the full cross-coupling structure. Compared to sparse identification of nonlinear dynamics~(SINDy) and its variants~\cite{brunton2016discovering,champion2019data,rudy2017data,li2026interpretable}, which build candidate libraries of spatial derivative terms and perform sparse regression directly on the field data, our approach benefits from the noise-filtering effect of operator learning. The regularized manifold optimization absorbs measurement noise before the PDE coefficients are extracted via spatial moments, rather than requiring differentiation of noisy fields. Compared to deep learning models, our framework gains interpretability, where every extracted quantity has a direct physical meaning corresponding to self-assembly process.

Our framework is extensible along several directions. Generalizing from two-dimensional imaging to three-dimensional electron tomography data would require replacing the 2D spatial grid with a 3D voxel mesh and the $5$-point stencil with a $7$-point~(or even larger) 3D stencil~\cite{jenkinson2022multimode,midgley2009electron,craig2026continuous,wells2008confocal}. The manifold optimization and moment extraction procedures carry over without conceptual modification, though the computational cost would increase. Systems with more complex order parameters, such as chiral, polar, or multi-component assemblies, can be accommodated by expanding the field vector $X$ to include additional order parameter channels, with the block structure of $A_{t}$ growing accordingly. Because our framework provides real-time estimates of transport coefficients as the assembly proceeds, these estimates could in principle serve as feedback signals to adjust experimental conditions and steer the system toward desired structural outcomes.

More broadly, this toolkit is applicable wherever experimental spatiotemporal field data is available, from active matter systems and biological pattern formation to materials processing and diffusion-reaction dynamics~\cite{serra2023defect}, and offers a principled route from observational data to interpretable physics, as well as design rules to program desired behavior and functionality into self assembly systems. 

\begin{acknowledgments}
The authors are grateful for the financial support from National Science Foundation (NSF) and specifically for grant \#2243104, Center for Complex Particle Systems (COMPASS), grant \#2317423.
\end{acknowledgments}

\bibliography{sn-bibliography}

\end{document}